\documentclass[10pt,letterpaper]{article}

\usepackage[top=0.85in,left=2.75in,footskip=0.75in]{geometry}
\usepackage{amsmath,amssymb}
\usepackage{changepage}
\usepackage[utf8x]{inputenc}
\usepackage{textcomp,marvosym}
\usepackage{cite}
\usepackage{nameref,hyperref}
\usepackage[right]{lineno}
\usepackage{microtype}
\DisableLigatures[f]{encoding = *, family = * }
\usepackage[table]{xcolor}
\usepackage{array}
\newcolumntype{+}{!{\vrule width 2pt}}
\newlength\savedwidth

\raggedright
\usepackage[aboveskip=1pt,labelfont=bf,labelsep=period,justification=raggedright,singlelinecheck=off]{caption}

\makeatletter
\renewcommand{\@biblabel}[1]{\quad#1.}
\makeatother

\usepackage{lastpage,fancyhdr,graphicx}
\usepackage{epstopdf}
\fancyheadoffset[L]{2.25in}
\fancyfootoffset[L]{2.25in}
\begin{document}

\vspace*{0.2in}
\begin{flushleft}
{\Large
\textbf\newline{
Beyond ranking nodes: Predicting epidemic outbreak sizes by network centralities
}
}
\newline
\\
Doina Bucur\textsuperscript{1*},
Petter Holme\textsuperscript{2},
\\
\bigskip
\textbf{1} University of Twente, The Netherlands
\\
\textbf{2} 
Tokyo Tech World Research Hub Initiative (WRHI), Institute of Innovative Research, Tokyo Institute of Technology, Yokohama, Japan
\bigskip

* d.bucur@utwente.nl

\end{flushleft}

\section*{Abstract}
Identifying important nodes for disease spreading is a central topic in network epidemiology. We investigate how well the position of a node, characterized by standard network measures, can predict its epidemiological importance in any graph of a given number of nodes. This is in contrast to other studies that deal with the easier prediction problem of ranking nodes by their epidemic importance in given graphs. As a benchmark for epidemic importance, we calculate the exact expected outbreak size given a node as the source. We study exhaustively all graphs of a given size, so do not restrict ourselves to certain generative models for graphs, nor to graph data sets. Due to the large number of possible nonisomorphic graphs of a fixed size, we are limited to ten-node graphs. We find that combinations of two or more centralities are predictive ($R^2$ scores of $0.91$ or higher) even for the most difficult parameter values of the epidemic simulation. Typically, these successful combinations include one normalized spectral centrality (such as PageRank or Katz centrality) and one measure that is sensitive to the number of edges in the graph.
 
\section*{Author summary}

A central challenge in network epidemiology is to find nodes that are important for disease spreading. Usually, one starts from a certain graph, and tries to rank the nodes in a way that correlates as strongly as possible with measures of importance estimated using simulations of outbreaks. A more challenging prediction task, and the one we take, is to ask if one can guess, from measures of the network structure alone, the values of quantities describing the outbreak. Having this predictive power is important: one can then target nodes that are more important than a certain threshold, rather than just a top fraction of nodes. By exhaustively studying all small graphs, we show that such prediction is possible to achieve with high accuracy by combining standard network measures.

\section*{Introduction}

Infectious diseases are still a major burden to global health. To mitigate them is of great societal value, and a cause to which theoretical modeling can be of help. Theoretical epidemiology has developed several core concepts that are guiding medical epidemiologists and public-health policy makers, including: epidemic thresholds, herd immunity, and the basic reproductive number~\cite{Anderson1991book,hethcote2000mathematics,giesecke}. There are a multitude of theoretical approaches to understanding the spreading of infections in populations---some more mathematical, some more computational. Our work models the underlying contact structure upon which the disease spreads as a network. This approach, \textit{network epidemiology}~\cite{KissMillerSimon2017book,Pastorsatorras2015RevModPhys}, is an emerging area with good prospects of improving epidemic forecasting~\cite{Colizza2015} and interventions~\cite{aiello}.

A common assumption of network epidemiology, and one we take, is that the disease spreads over a network that is evolving much slower than the disease outbreak. In this case, the propagation of an outbreak can be modeled by a \textit{compartmental model}. Such a model divides the population into states with respect to the disease such as: \emph{susceptible} (S; who can get the disease), \emph{infectious} (I; who can infect susceptibles), and \emph{recovered} (R; who neither can get the disease, nor can infect others), and assigns transition rules between the classes. With this setup, one of the most common research questions is that of finding which network characteristics predict the importance of a node with respect to the disease spreading~\cite{vaccination,lu2016vital,Pastorsatorras2015RevModPhys}. More precisely, these authors seek network structural measures that rank nodes in the same order as some quantity describing their importance with respect to the disease spreading~\cite{Holme_2004,vsikic2013epidemic,bauer2012identifying}. Some authors investigate the predictive power of such ``centrality measures''~\cite{arruda,ames2011using,miller:vacc,rushmore} (a term we use, although it is somewhat ambiguous), but none as far as we know study combinations of centralities.

For interventions (vaccination, quarantine, pre-exposure prophylaxis, etc.)\ based on network measures to become useful to public health practitioners, there are several hurdles to overcome. A network obtained by, e.g., contact tracing~\cite{giesecke} will be both noisy and incomplete. Some studies have investigated the robustness to noise of network measures to identify importance~\cite{smieszek2013low,borgatti2006robustness,genois2018can}, and other studies have investigated how incomplete data based on questionnaires and observations are~\cite{wilder2017uncharted}. In practice, one cannot expect a contact network to be either completely accurate or possible to map out. (Some types of networks are more controllable than others: a network of animal trade is one example relevant to epidemic modeling, where one can both credibly infer the links and trust them.) The problem of how to estimate centrality measures from incomplete data is an emerging research topic~\cite{zhang2015estimating,ruggeri}. Ultimately, one would like to combine such results with our approach to construct more practical methods of inferring important nodes. For this paper, however, to facilitate comparison with existing studies (such as Refs.~\cite{smieszek2013low,borgatti2006robustness,genois2018can}) we stick to well-studied and established centrality measures.

\begin{figure}[bth]
 \includegraphics[width=0.85\columnwidth]{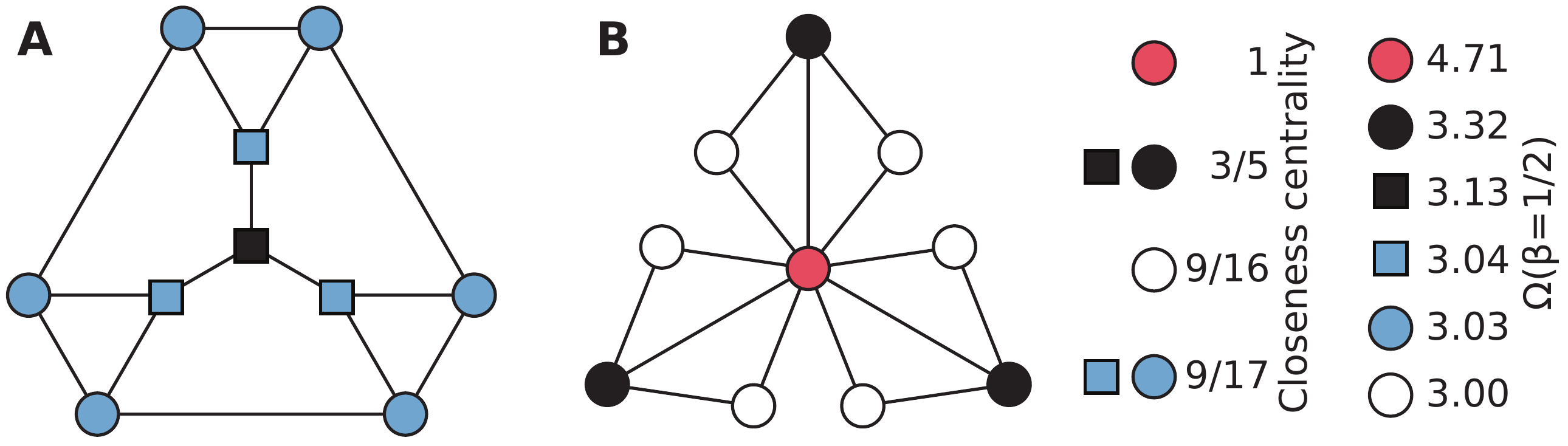}
 
 \caption{{\bf Comparing nodes in different graphs by closeness centrality and $\Omega$.} The black nodes in the two graphs all have closeness $3/5$, which ranks them as the most important node in panel A but of intermediate importance in panel B. The value $3/5$ is thus insufficient for ranking the nodes importance. Closeness manages to rank the nodes within each graph correctly with respect to $\Omega$ for the infection rate $\beta=1/2$ (except it does not split the blue nodes of the graph in panel A), but ranks the white nodes of panel B too high in both graphs together.}
 \label{fig:ill}
\end{figure}

Yet another issue is how to compare network predictors of importance from different data sets. In sparser networks, an outbreak needs less disease control to be contained, so, say, the third highest ranked individual would, in absolute terms, not be as important as the third highest ranked one in a denser network. Even if networks have the same number of nodes and edges this kind of effect can occur. In Fig.~\ref{fig:ill}, we illustrate the different aspects of using a network centrality (in this figure, closeness centrality) to predict an importance measure based on epidemic models---in our case the expected \textit{outbreak size} $\Omega$ (sometimes called \textit{attack rate}) if any node is the seed of the infection. Our paper explores the raw value of centralities such as closeness in predicting the importance of nodes with respect to disease spreading.

Ideally one would not like a ranking of nodes for a specific network, but an \emph{absolute} way to compare nodes across networks. If an application needs to target all nodes more important than a threshold, then a ranking of nodes per network would not suffice. To properly address the question of how the values of structural predictors of network importance can predict the outbreak size in arbitrary graphs, we cannot restrict ourselves to networks generated by a random model. If we did, we would not be able to say whether our results are consequences of the model, or of the inherent constraints placed on the disease spreading by the network. Instead of sampling graphs from a network model, we study all graphs. A drawback of this approach is that, since the number of graphs of a certain number of nodes grows very fast, we will be restricted to small graphs up to size $N=10$. Although we ultimately want to generalize our results to large graphs, there are many advantages to studying only small graphs. First, one can use slower, exact algorithms to determine the outbreak size~\cite{3faces}. This is important since often, the numerical difference in the objective importance measure are too small to be separated in stochastic simulations even with large averages. Second, we do not have to restrict ourselves to network models. Because the graphs are small, we can scan them exhaustively, and thus identify innate effects of the underlying contact structure. Third, many scaling properties of graphs hold already for small graphs~\cite{BARABASI1999173,Ozana_2001}. Fourth, there are small networks which are relevant to medical epidemiology. For example, networks of farms connected by animal transport are deliberately kept small and disconnected to prevent the introduction of disease~\cite{bajardi,dawson2015epidemic}. These could be modeled by metapopulation dynamics~\cite{chowell2016mathematical}, or (as we do) the standard compartmental models with nodes representing the farms. Fifth, large networks could be reduced to smaller ones by community detection~\cite{santo}. This is similar to point four but without the meta information of what individual (animal) then belongs to what group (farm).

The outline of our method is to calculate the exact outbreak size $\Omega_i$ given that a disease starts at a node $i$. This is usually called the \emph{influence maximization} problem~\cite{kempe2003maximizing} or sometimes the problem to identify \emph{super spreaders}~\cite{radicchi2017fundamental} (but note that ``super spreaders'' has a different definition in the medical literature~\cite{giesecke}). Note that there are also other notions of what characterizes important individuals with respect to outbreak characteristics (such as the effect of vaccinating them) leading to somewhat different results~\cite{3faces}. Assuming the standard, Markovian Susceptible--Infectious--Recovered (SIR) model, we calculate $\Omega_i$ exactly for every node in every connected graph of $6\leq N\leq 10$ nodes. Then we ask how well standard networks predictors of node importance (such as degree or betweenness centrality)~\cite{arruda}, and particularly combinations of these, can predict $\Omega_i$. We follow a statistical learning approach: we split the data into training and validation parts; we use standard supervised learning algorithms (Random Forest and Support Vector Machine regression); we use the coefficient of determination as a performance metric and permutation tests with ten-fold cross validation for significance testing.

\section*{Methods}

\subsection*{Computing $\Omega$ exactly in the SIR model} 

\begin{figure}[tb]
 \includegraphics[width=.95\textwidth]{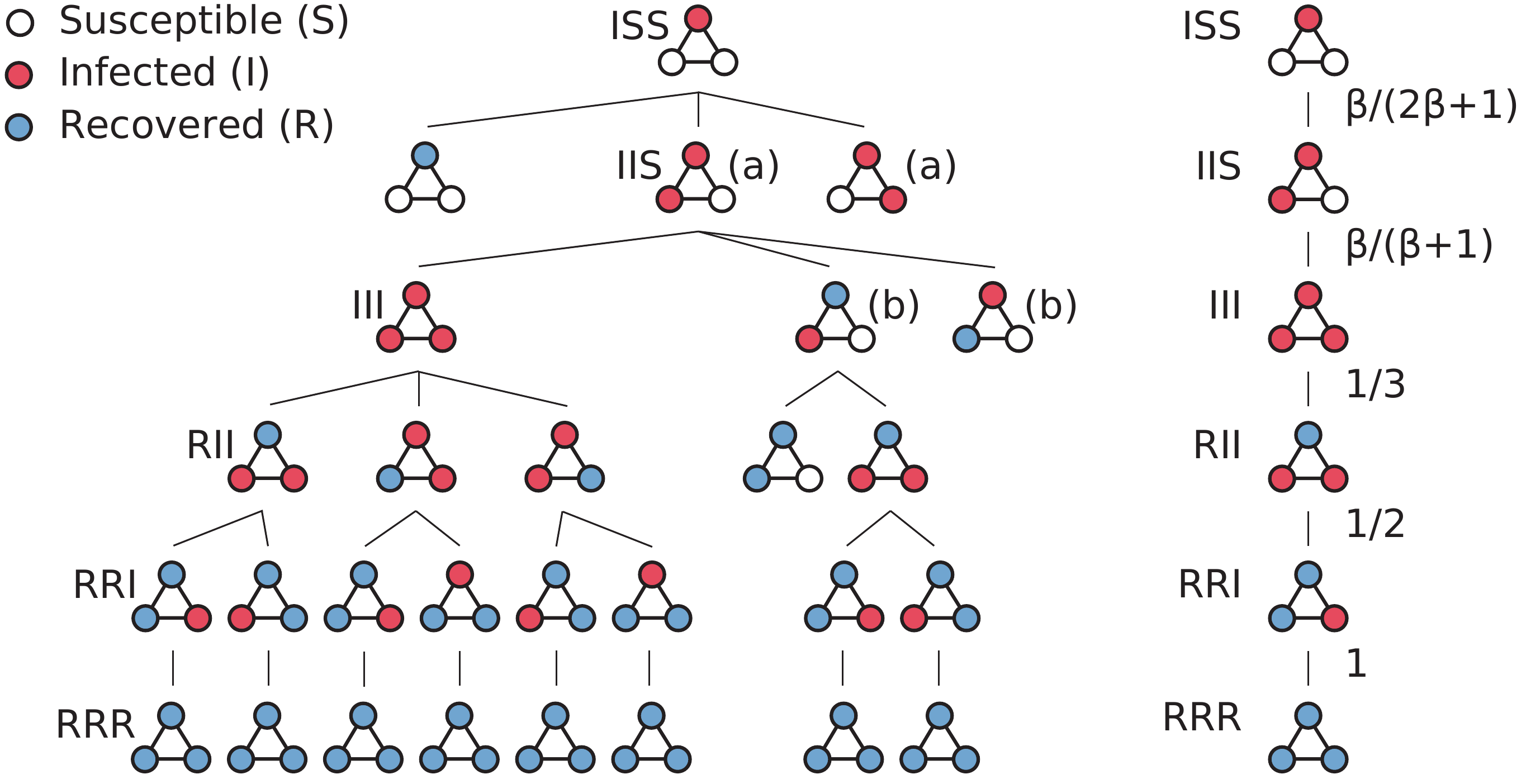}
 
 \caption{{\bf The unfolding of the SIR configuration tree.} (left) For the triangle graph with one initial infected node, the outbreak is a tree of configurations. The subtrees whose root nodes are labeled (a) unfold symmetrically (only one is shown); the same for (b). (right) For a path in the tree, the transition probabilities are shown.}
 \label{fig:SIR}
\end{figure}

In the SIR model, at any given time, each of the $N$ nodes of an undirected graph $G$ is in one of the (above mentioned) states: S, I, or R. Susceptible and infected nodes may transition into other states via two types of events:
\begin{description}
 \item[Infection events] A susceptible node connected to an infected node becomes infected at a rate of $\beta$ infection events per time unit.
 \item[Recovery events] An infected node recovers at a rate of $\nu$ recovery events per time unit. We measure time in units of $\frac{1}{\nu}$, so that the recovery rate becomes $\nu=1$, and the SIR model has only $\beta$ as a parameter.
\end{description}
At any given time during an outbreak of the Markovian SIR model, the system is fully determined by its \emph{configuration} $C$ of susceptible, infectious and recovered nodes. The probability of the next event being an infection event is
\begin{equation}
P_{\rm infection}=\frac{\beta M_{SI}}{\beta M_{SI}+N_{I}} ,
\end{equation}
and that of the next event being a recovery event is
\begin{equation}
P_{\rm recovery}=\frac{N_{I}}{\beta M_{SI}+N_{I}},
\end{equation}
where $M_{SI}$ is the number of edges between nodes in the states S and I, and $N_{I}$ the number of infected nodes~\cite{3faces}. By repeatedly applying these two rules to an initial graph of all nodes susceptible except one (the infection seed), we can calculate the exact probability of every configuration. To illustrate this, we show the run of the SIR model over the triangle graph, as a branching tree of configurations rooted in the initial configuration ISS (Fig.~\ref{fig:SIR} left). The right panel of Fig.~\ref{fig:SIR} shows the transition probabilities for one possible run of the outbreak. The probability of reaching any configuration $C$ of any run is simply the product of all the transition probabilities on the tree path to $C$. The probabilities of the final configurations (those without any infected nodes) give the expected outbreak size $\Omega$, which is our key quantity describing the potential severity of the epidemic outbreak. Note that $\Omega$ is a deterministic quantity even though the SIR model is a stochastic process.

The \emph{expected outbreak size} $\Omega$ is the expected number of nodes in $G$ which have been infected during the outbreak. Since, for any infected node, a recovery is eventually guaranteed, this is equivalent to the expected number of recovered nodes, denoted $N_{R}$. Computing the exact value for $\Omega$, given $\beta$, requires the unfolding of the complete tree of configurations; $\Omega$ is then the sum of $N_{R}$ across all final configurations, weighted by the probability of reaching each final configuration.

A number of optimizations are possible when computing $\Omega$. To avoid the exploration of identical configurations multiple times, the tree is explored with a breadth-first strategy. Since the model is Markovian, whenever two identical configurations are reached, they can be merged by summing their probabilities, and are only explored once. For example, configuration III in Fig.~\ref{fig:SIR} is reachable via two paths, in the subtrees marked there with (a). Also, when equivalent nodes in the same graph are the initial infection sites, the computation is only done once. 

We collect exact numerical results for $\Omega$ for the nine $\beta$ values in a geometric sequence with common ratio 2, between the values $1/16$ and 16. The computation for all values of $\beta$ is done in the same exploration run. Across graphs of $N=10$ nodes and with C++ code, the average runtime on a 3.1-GHZ CPU is 0.2 seconds per graph.

\subsection*{All nonisomorphic graphs as a graph model}

\begin{figure}[tb]

 \includegraphics[width=\textwidth]{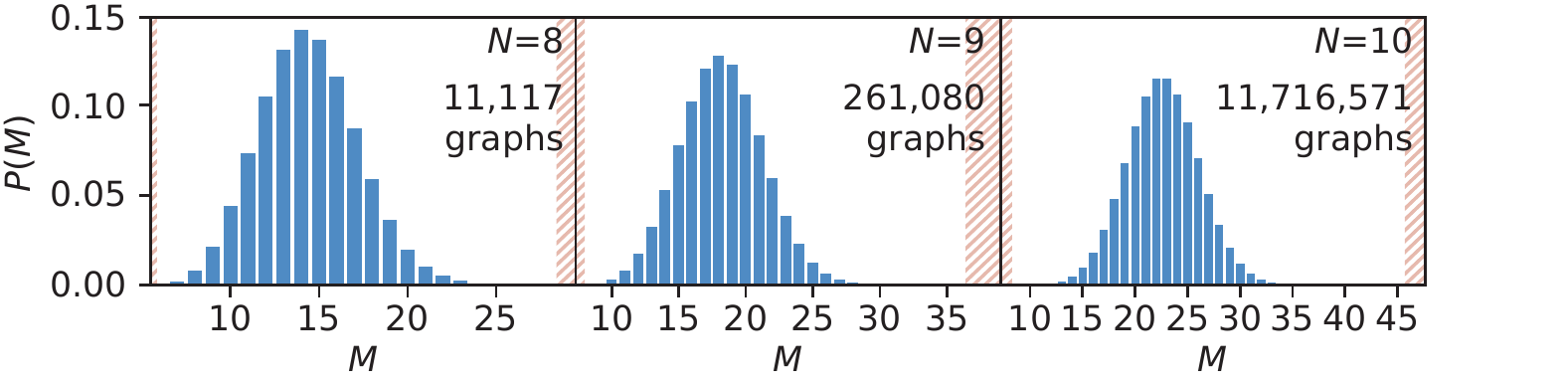}

 \caption{{\bf Small nonisomorphic graphs.} The discrete probability distribution for the number of edges $M$ across all nonisomorphic, connected, undirected graphs of $8 \le N \le 10 $. The shaded areas mark values outside of the bounds of $M$ ($N-1\leq M\leq N(N-1)/2$).}
 \label{fig:edge_distributions}
\end{figure}

We generated all nonisomorphic, connected, simple undirected graphs of $N \le 10$ nodes  with the tool \texttt{geng}~\cite{mckay2014practical}. There are 112 graphs of six nodes, but 11.7 million graphs of ten nodes. The graphs have similar shapes of their discrete probability distributions for the number of edges $M$ (and these are shown in Fig.~\ref{fig:edge_distributions}).

The set of nonisomorphic graphs is not a graph model \textit{per se}, because it does not assign a probability to each graph. Studying it is still interesting since it allows us to explore the full range of possibilities. One can, however, turn this set into a model by assigning an equal weight to every graph. For our statistical analysis this is what we do. In other words, we use a model that enables us to understand the inherent features of connected graphs. But most importantly, we can scan all members of this model and thus cover its extreme features, something that would not be possible with a model that one would need to sample stochastically. 

For every graph size $N \le 10$, we form a data set. A record (or row) in this data set describes any node $i$ from any graph $G$ via the following data columns: an identifier for $G$, an identifier for $i$, the values of centrality measures for node $i$, and the exact numerical results for the outbreak size $\Omega$ (using $i$ as the only infection seed) for the nine $\beta$ steps between $1/16$ and 16. A graph $G$ is represented in the data set $N$ times, in records describing the importance and extent of outbreak for each of the graph's nodes. Thus, the number of records in the data set for a given graph size $N$ is $N$ times the number of nonisomorphic, connected, undirected graphs of that size; this means 117 million records for $N=10$.

\subsection*{Centrality measures}

Any of the nodes in a graph may be the one starting an outbreak. As descriptive features for the nodes, we use seven standard network measures---most of the usually branded as centrality measures---which capture different aspects of a node's importance in an undirected, connected graph~\cite{newman2010networks,arruda}. These are defined in Table~\ref{tab:centralities}. PageRank and Katz centrality take a parameter $\alpha$: for PageRank, $\alpha=0.85$ (the ``damping factor''), while for Katz centralities, $\alpha=0.1$ (the ``attenuation factor'').

\begin{table}[tb]
\begin{adjustwidth}{-2.25in}{0in}

 \caption{{\bf Centrality measures} In matrix notation:
 $\mathbf{x}$ is the vector of node centralities, 
 $\mathbf{A}$ is the graph adjacency matrix, 
 $\lambda_1$ is the largest eigenvalue of $\mathbf{A}$, 
 $\mathbf{D}$ is the degree matrix (the diagonal matrix of node degrees), 
 $\mathbf{1}$ is the vector of ones, and 
 $\mathbf{I}$ is the identity matrix (the diagonal matrix of ones). 
 Other notation: 
 $d_{ij}$ is the number of edges on the shortest path between nodes $i$ and $j$,
 $\sigma_{jk}$ is the number of shortest paths between nodes $j$ and $k$,
 $\sigma_{jk|i}$ is the number of shortest path between nodes $j$ and $k$ which pass through $i$.}
 \label{tab:centralities}
 
 \begin{tabular}{lll}
 {\bf Centrality} & {\bf Definition}\\ \hline
 Degree centrality & $\mathbf{x} = \mathbf{A}\mathbf{D}^{-1}\mathbf{x}$ \\
 Eigenvector centrality & $\mathbf{x} = \lambda_1^{-1}\mathbf{A}\mathbf{x}$ \\
 PageRank & $\mathbf{x} = \mathbf{D} \left(\mathbf{D}-\alpha \mathbf{A}\right)^{-1}\mathbf{1}$ \\
 Katz centrality & $\mathbf{x} = \left(\mathbf{I}-\alpha \mathbf{A}\right)^{-1}\mathbf{1}$ \\
 \end{tabular}
 \hspace{1cm}
 \begin{tabular}{lll}
 {\bf Centrality} & {\bf Definition}\\ \hline
 Closeness centrality & $C_i = (N-1) / \sum_{j} d_{ij}$ \\
 Betweenness centrality & $C_i = \sum_{j,k} \sigma_{jk|i} / \sigma_{jk}$ \\
 Coreness & $C_i = $ largest $k$ so $i$ is in a $k$-core \\
 & \\
 \end{tabular}

\end{adjustwidth}
\end{table}

All network measures are normalized to the $[0,1]$ range. For the degree centrality, the node degrees are divided by the maximum degree $N-1$. Similarly, the closeness, betweenness, and coreness centralities are normalized so that the maximum value is one. The eigenvector- and Katz centrality are normalized by the Euclidean length (or 2-norm) of the vector of node centralities $\mathbf{x}$, while the PageRank centralities in $\mathbf{x}$ are normalized so they sum to one.

We use the edge density $M/N$ (which is equal to half of the average degree) as an eighth predictor. It is also normalized to the $[0,1]$ range.

\subsection*{Supervised learning for predicting $\Omega$}

In order to understand the fundamental ability of the centrality measures in graphs of size $N$ to predict the target variable $\Omega (\beta)$, all small combinations of centralities are tried out as predictor variables (or features). We thus set up the following experiments: for every $6 \le N\le 10$ ($6$ being the smallest graph size which allows sufficient data), for every $1/16 \le \beta \le 16$ (with nine values for $\beta$ in a geometric sequence), and for every combination of centralities, a regression analysis is run using the relevant data columns in the complete data set for $N$. Each regression analysis trains, tunes, and cross-validates a statistical model on a development-fraction of the data, and tests the tuned model on the remaining test data. The test results are then reported, and some of the resulting models are also visualized.

\paragraph{Unique target-predictor data records.} Assume a given $N$ and $\beta$. All single centrality measures, and all combinations of two and three of these are selected as predictors for the same target $\Omega(\beta)$ in independent analyses. A fraction of these selected data records are exact duplicates; this happens primarily for records describing automorphically equivalent nodes. All duplicates are removed from the data prior to the regression analysis, so that none of the test data is identical to any training data.

\paragraph{The train-test split and the learning curve.} It is not clear a priori how to split the data set into \emph{development data} (for training and validation) and \emph{test data}. Particularly when the data is abundant (the case when $N$ is large), the development data need only be as large as necessary. In other cases, the development data should instead be larger, to avoid learning a high-variance (or overfitted) statistical model. For this, regardless of the particular regression algorithm used, the size of the development data is treated as a hyperparameter, and is tuned. Ten data sizes are selected on a linear scale up to a maximum size. Then, a regressor is trained and cross-validated using ten-fold cross-validation on randomly sampled training data of each required data size, and the training and validation performance are plotted against the data size. A suitable development data size is that at which the validation curve (a) is close to the training curve, and (b) levels off, so that increasing the data size brings no further advantage. We illustrate the convergence of the training data in Fig.~S1. Other learning curves  show a similar convergence.

While 75\% of the $N=6$ data set (around 400 data points) is needed as training data, 5\% is sufficient at $N=10$ (around 5 million data points, varying  with each target-predictor combination). All remaining data is used as test data. 

\paragraph{Regression algorithms and hyperparameter tuning.} We use two algorithms for statistical learning: Random Forest Regression (RFR) and Support Vector Machine Regression (SVR) 
and their implementations from the Scikit-learn machine-learning library~\cite{scikit-learn}. These statistical models are different in design. While SVR solves an optimization problem, RFR is an ensemble of weak learners (decision trees), each trained independently with a greedy heuristic using a bootstrap sample from the training data (see Ref.~\cite{friedman2001elements} for algorithmic details).

Both algorithms are able to learn nonlinear relationships between multiple predictors and a target variable, and have hyperparameters which are themselves trained using a grid search with cross-validation. For RFR, the hyperparameters are (a) the number of decision trees (up to 20), and (b) the minimum number of samples required to be at a leaf node (tuned between 1\% and 0.01\% of the size of the training set); the latter also helps to control overfitting~\cite{friedman2001elements}. For SVR, the hyperparameters are (a) the type of kernel (either linear or a radial basis function, RBF), (b) the regularization parameter $C$ (tuned between 0.1 and 100)~\cite{scikit-learn}. In all cases, SVR models are configured with an automatically scaled kernel coefficient $\gamma$, and a distance of estimation at which no penalty is given in the training loss function $\epsilon=0.01$~\cite{scikit-learn}. 

In our study, only small combinations of two or three predictors are used; some of these predictors may be correlated (for example, a node's degree centrality and a spectral measure), but we still want to study their combined predictive power. Both algorithms yield stable prediction values in the presence of correlated predictors. In particular, RFR is designed to be resilient even with strongly correlated variables: it averages the predictions of many independently trained decision trees, which acts as a stabilizer allowing two strongly correlated variables to both be important in the model~\cite{friedman2001elements}.

RFR scales best computationally with an increasing size of the training data. On the other hand, unlike RFR, SVR models with either a linear or an RBF kernel obtain a smooth, continuous regression landscape, which, when visualized, is easily interpretable. Both regressors achieve similar performance scores on the data in this study. In the Results section, we report the performance scores of the RFR models, which are the most efficient to train among all. When visualizing the statistical models obtained, we use instead the more interpretable SVR models.

\paragraph{The performance metric $R^2$.} The \emph{coefficient of determination} $R^2$ serves as the scoring function for any regressor. This is the fraction of the variance in the target that was predicted correctly, and has the expression $1-S_\text{res}/S_\text{tot}$, where $S_\text{res}$ is the residual sum of squares (or the distance between the test data and the estimation) and $S_\text{tot}$ is the total sum of squares (of the target data points to the target mean). A perfect model has $R^2=1$. A constant model which predicts the target mean will score $R^2=0$; arbitrarily large negative values are possible. 

\paragraph{Significance tests.} We also further evaluate the significance of the regression with permutation-based p-values. The target values are permuted so that any structural dependency between target and predictors is lost; then, a ten-fold cross-validation is performed on the development data, with each fold trained on 100 permutations. This tests the following null hypothesis: the predictor data and the target data are independent, so no relationship between them can be significant~\cite{ojala2010permutation}. We always obtain the minimum p-value possible, which rejects the null hypothesis, and confirms that a true dependency is discovered.

\section*{Results}

\subsection*{Examples}

\begin{figure}[tb]
\begin{adjustwidth}{-1.4in}{0in}

  \includegraphics[width=1.2\textwidth]{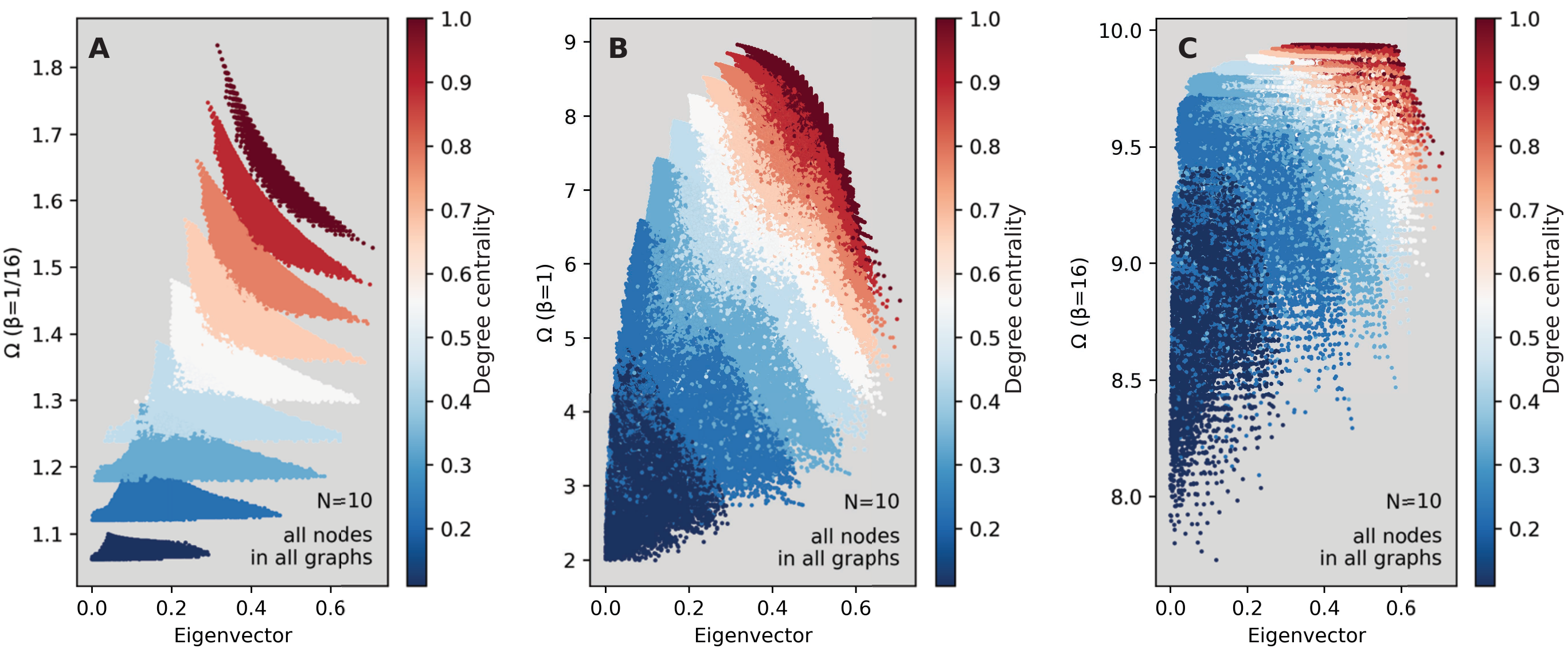}

 \caption{{\bf How the expected outbreak size $\Omega$ varies with a spectral centrality measure and the node degree across all small graphs.} Panels A, B and C show $\Omega$ for any node starting an outbreak, in any graph of size $N=10$, against the eigenvector centrality of that node. Panel A shows data for $\beta=1/16$; B for $\beta=1$ and C for $\beta=16$. The color denotes the degree centrality.}
 \label{fig:eigen_vs_omega_scatterplot_degree}

\end{adjustwidth}
\end{figure}

We start our exhibition of results by studying an example---the raw scatter plots of $\Omega$ as a function of the eigenvector centrality, in Fig.~\ref{fig:eigen_vs_omega_scatterplot_degree}A--C. Every point in these figures corresponds to one node in one ten-node graph. The color represents the degree centrality of every node. In panel A---corresponding to a very small transmission rate ($\beta=1/16$)---we can see the nodes of different degrees grouping together into (partly overlapping) clusters, with the clusters corresponding to higher values for the degree centrality also having comparatively higher $\Omega$ values. On the other hand, the value of the eigenvector centrality does not correlate strongly with $\Omega$, and the nodes with the highest eigenvector centrality do not also have the highest $\Omega$ value. Note that, even though e.g.\ Fig.~\ref{fig:eigen_vs_omega_scatterplot_degree}B looks like generated by a random process, it is not. Everything comes from the restriction of graphs to be simple and of ten nodes.

Consequently, for $\beta$ this low, the value of the degree is expected to be much more predictive of that outbreak size than the eigenvector centrality: knowing the value of the degree leads to being able to estimate $\Omega$ within a small interval. This is easy to understand, since the probability of any outbreak is small and decaying fast with the size of the outbreak~\cite{3faces}: if the outbreak does not die immediately, only the neighbors of the seed node are infected. The more neighbors, the larger the outbreak---hence the tidy clustering by degree at this small value of $\beta$. 

As $\beta$ increases---panels B and C in Fig.~\ref{fig:eigen_vs_omega_scatterplot_degree}---the clusters defined by the degree merge. When $\beta=1$ (panel B), the eigenvector centrality becomes a slightly better predictor of $\Omega$ than in panel A, but still far from good. At this value of $\beta$, neither the degree, not the eigenvector centrality appear to be good predictors when considered individually, but their combination is promising: knowing both may lead to a good estimation of $\Omega$. Furthermore, note that for the intermediate value $\beta=1$, the range of $\Omega$ values (around $7$) is much larger than panels $A$ (around $0.8$) and $C$ (around $2$) which illustrates the non-linearity of the SIR model even in small networks. As well-known~\cite{Pastorsatorras2015RevModPhys}, when $N\to\infty$ such non-linearities will sharpen to a threshold separating one phase where the disease can spread to a finite fraction of the population and one phase where the outbreaks will always be small.

\begin{figure}[tb]
\begin{adjustwidth}{-1.4in}{0in}

  \includegraphics[width=1.2\textwidth]{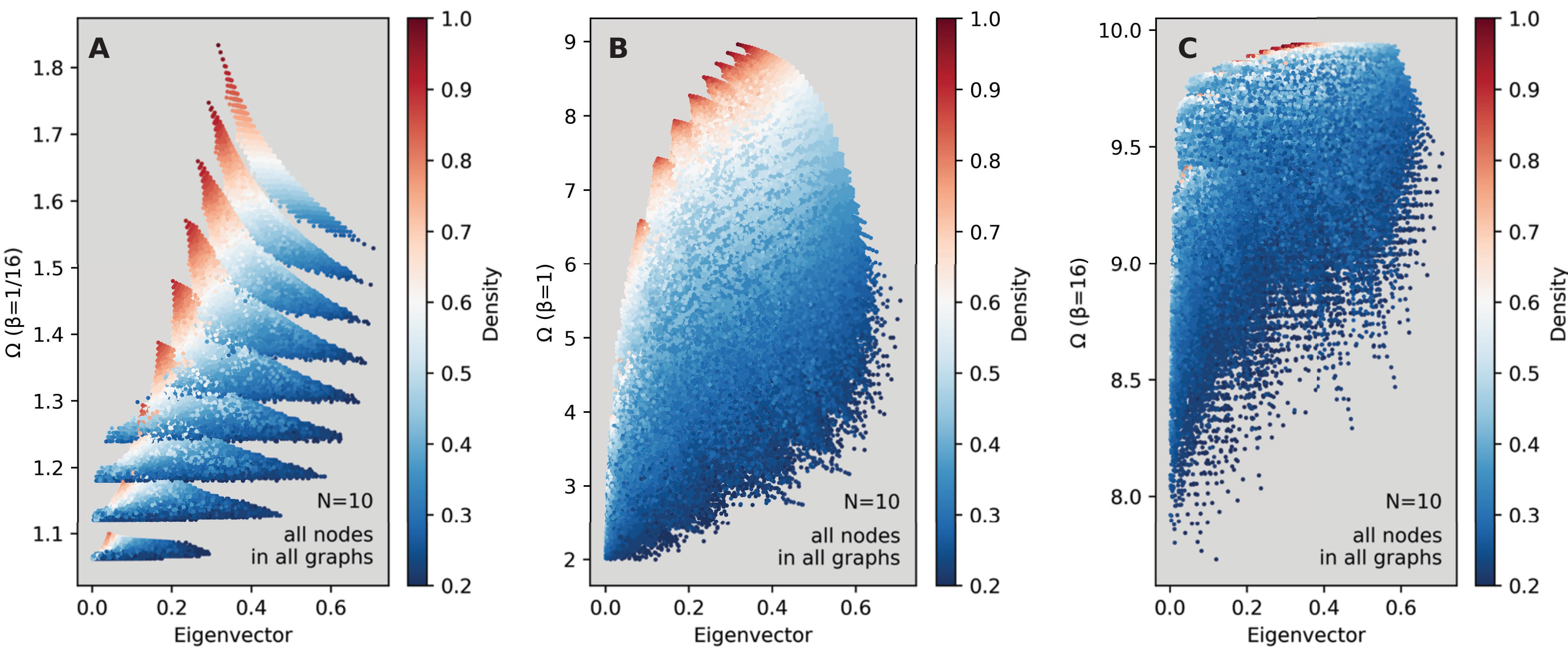}

 \caption{{\bf How the expected outbreak size $\Omega$ varies with a centrality measure and the edge density across all small graphs.} As Fig.~\ref{fig:eigen_vs_omega_scatterplot_degree}, except that the color here denotes the network density.}
 \label{fig:eigen_vs_omega_scatterplot_density}

\end{adjustwidth}
\end{figure}

The edge density of the networks also gives interesting scatterplot patterns. Fig.~\ref{fig:eigen_vs_omega_scatterplot_density} shows the same example as Fig.~\ref{fig:eigen_vs_omega_scatterplot_degree}, except here the colors show the edge density of the network from which each node originated. This figure demonstrates the secondary effect of connections beyond the seed node---in denser networks (redder nodes in the figure) there are more opportunities of tertiary (and further, higher-order) infections, so the clusters of nodes which have similar density values now have a large vertical spread, do not correspond with the degree clusters, and tend to have low and medium eigenvector centrality values. When knowing both the value of the edge density, and that of the eigenvector centrality, one may be able to estimate $\Omega$ to within a small interval, at least for low and medium $\beta$ values. The corresponding plots for PageRank are shown in Figs.~S2 and S3. They show how PageRank separates clusters better and thereby outperforms eigenvector centrality as a predictor.

Note that the vast majority of nodes are a shade of blue in Fig.~\ref{fig:eigen_vs_omega_scatterplot_density}---cf.\ the probability distribution for the number of edges, in Fig.~\ref{fig:edge_distributions}---so that the scatter plots of panels B and C primarily look blue does not mean that the density of points at the red end of the color spectrum is higher. 

\begin{figure}[tb]

 \includegraphics[width=0.8\textwidth]{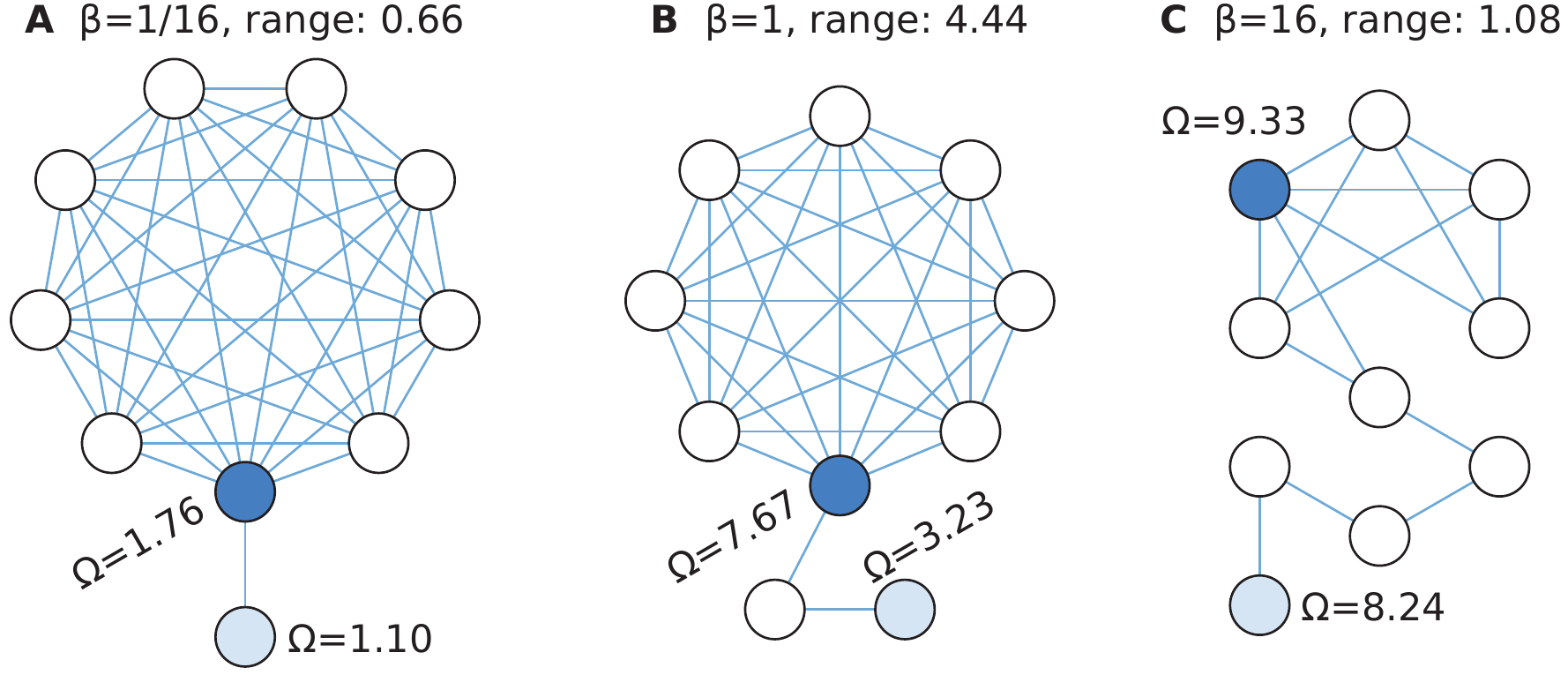}

 \caption{{\bf Example graphs with large $\Omega$ diversity.} Panel A shows the graph with largest range of $\Omega$ values for $\beta=1/16$; B for $\beta=1$ and C for $\beta=16$.}
 \label{fig:extreme_examples}

\end{figure}

Interestingly, the range of $\Omega$ within single networks is not much smaller than the entire range of $\Omega$-values (for all nodes in all networks). In Fig.~\ref{fig:extreme_examples} we show some networks with extreme ranges of $\Omega$. For all of these, the nodes of the highest $\Omega$ belong to a densely connected part of the networks (typically a clique), and the one with the smallest $\Omega$ is a degree-one node at the end of a chain-like protrusion from the dense cluster. Probably this description holds for all extreme examples, at least for small enough $\beta$.

\subsection*{Single-measure predictability}

\begin{figure}[tb]
\begin{adjustwidth}{-2.1in}{0in}

 \includegraphics[width=1.4\textwidth]{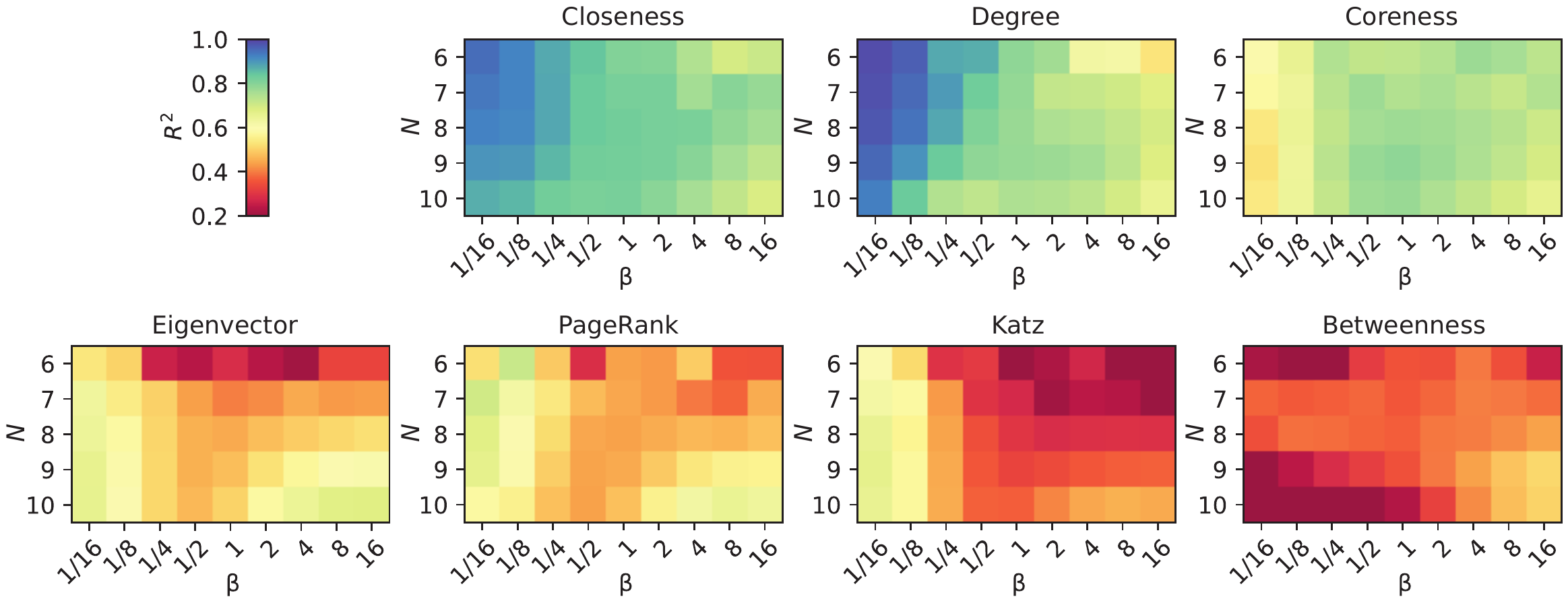}

 \caption{{\bf How predictive is a single node centrality?} The coefficient of determination $R^2$ when $\Omega(\beta)$ is estimated over graphs of $N$ nodes. The centralities appear in decreasing order of the minimum $R^2$ across $\beta$ values at $N=10$: 0.69 for closeness, 0.65 for degree, 0.54 for coreness, but only 0.12 for betweenness.}
 \label{fig:regression_summary-1centr}

\end{adjustwidth}
\end{figure}

In the previous section, we studied the relationship between the eigenvector centrality of nodes and the expected outbreak size if the nodes are the infection sources. In this section, we scale up to all seven centrality measures and all nine $\beta$ values we study. As a correlation measure, we use the coefficient of determination $R^2$ (see the Methods section). In Fig.~\ref{fig:regression_summary-1centr}, we plot heatmaps of the performance of our centralities as predictors for $\Omega$. First we note that this analysis confirms that the degree is a good predictor for small $\beta$, confirming an observation in Ref.~\cite{arruda}. Ref.~\cite{ames2011using} argues that the degree controls the disease spreading for both small and large $\beta$ (but not intermediate $\beta$); in our study it is less successful at large $\beta$. For medium and large $\beta$, closeness is the better network predictor.

The only measure fairing worse than the three spectral centralities (eigenvector centrality, Katz centrality and PageRank) is betweenness centrality. The rationale of the betweenness derives from an imagined dynamic system where packets are routed along shortest paths, which clearly is very far from the SIR model~\cite{centralities}. For example, being connected to a node that is very easily infected would make you easily infected. That recursive logic does not apply to the betweenness centrality. It does, however, apply to the spectral centralities, so why they perform worse than closeness, degree and coreness is harder to understand. Ref.~\cite{kitsak2010identification} promotes coreness as a importance predictor, so for medium and large $\beta$ we confirm that observation (but for low $\beta$, coreness is not performing very well). The spectral centralities can be motivated from random walk processes~\cite{newman2010networks}. These are less sensitive to parameter values compared to compartmental disease spreading models (they lack the threshold behavior of the latter). On the other hand, compartmental models far from the threshold are less sensitive to the network structure.

We cannot think of a quick explanation why closeness centrality has such high predictive power. It has been noted before~\cite{3faces} but is probably restricted to small graphs. Some authors have pointed out that closeness centrality becomes less useful for larger graphs~\cite{newman2010networks}. One argument is that the centrality of any node $i$ should be most dependent on nodes in the extended neighborhood $\Gamma_D(i)$ (i.e.\ the part of the network within a certain distance $D$ from $i$). However, for closeness centrality, the contribution of nodes in $\Gamma_D(i)$ goes to zero as $N$ increases. Making any change to $\Gamma_D(i)$ other than disconnecting $i$ from the bulk of the network will almost not change its closeness centrality for large enough networks. Our study, however, concerns small networks and in this realm, closeness centrality is apparently more useful.  

\begin{figure}
\begin{adjustwidth}{-2.1in}{0in}

 \includegraphics[width=1.4\textwidth]{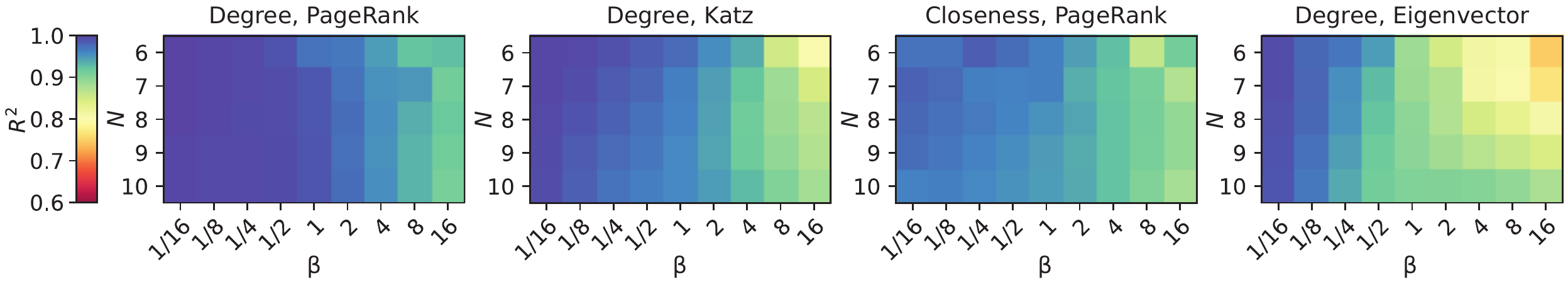}
 
 \caption{{\bf The most predictive pairs of node centralities.} The coefficient of determination $R^2$ when $\Omega(\beta)$ is estimated over graphs of $N$ nodes. The centrality pairs appear in decreasing order of the minimum $R^2$ across $\beta$ values at $N=10$: from 0.91 for degree and PageRank, to 0.88 for all three other combinations. }
 \label{fig:regression_summary-2centr}

\end{adjustwidth}
\end{figure}

\begin{figure}
\begin{adjustwidth}{-2.1in}{0in}
 \includegraphics[width=1.4\textwidth]{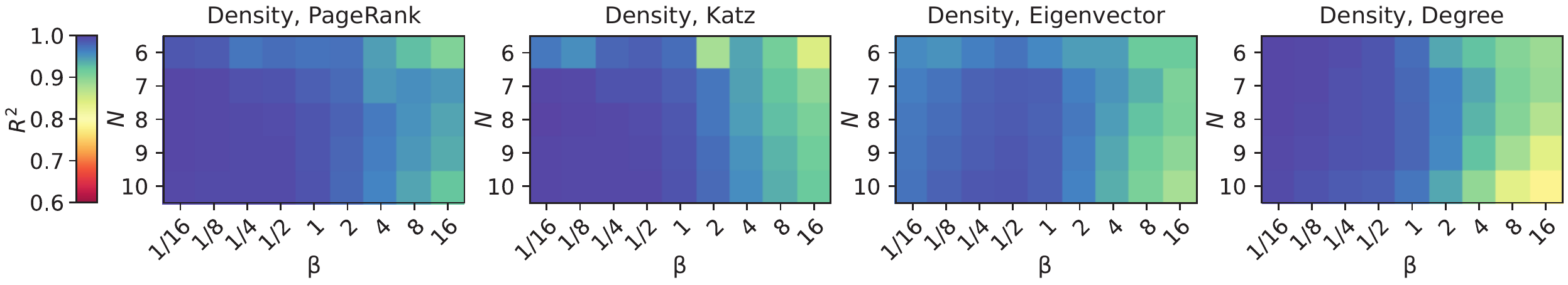}

 \caption{{\bf The predictability of one node metric and density (the number of edges in the graph of a node).} The coefficient of determination $R^2$ when $\Omega(\beta)$ is estimated over graphs of $N$ nodes similar to Fig.~\ref{fig:regression_summary-2centr}. The panels show the top four centralities in terms of the minimum $R^2$ value over all parameter combinations: $R^2=0.92$ for both PageRank and Katz, 0.88 for eigenvector centrality and 0.78 for degree.}
 \label{fig:regression_summary-density-1centr}

\end{adjustwidth}
\end{figure}

\begin{figure}
 \includegraphics[width=0.98\textwidth]{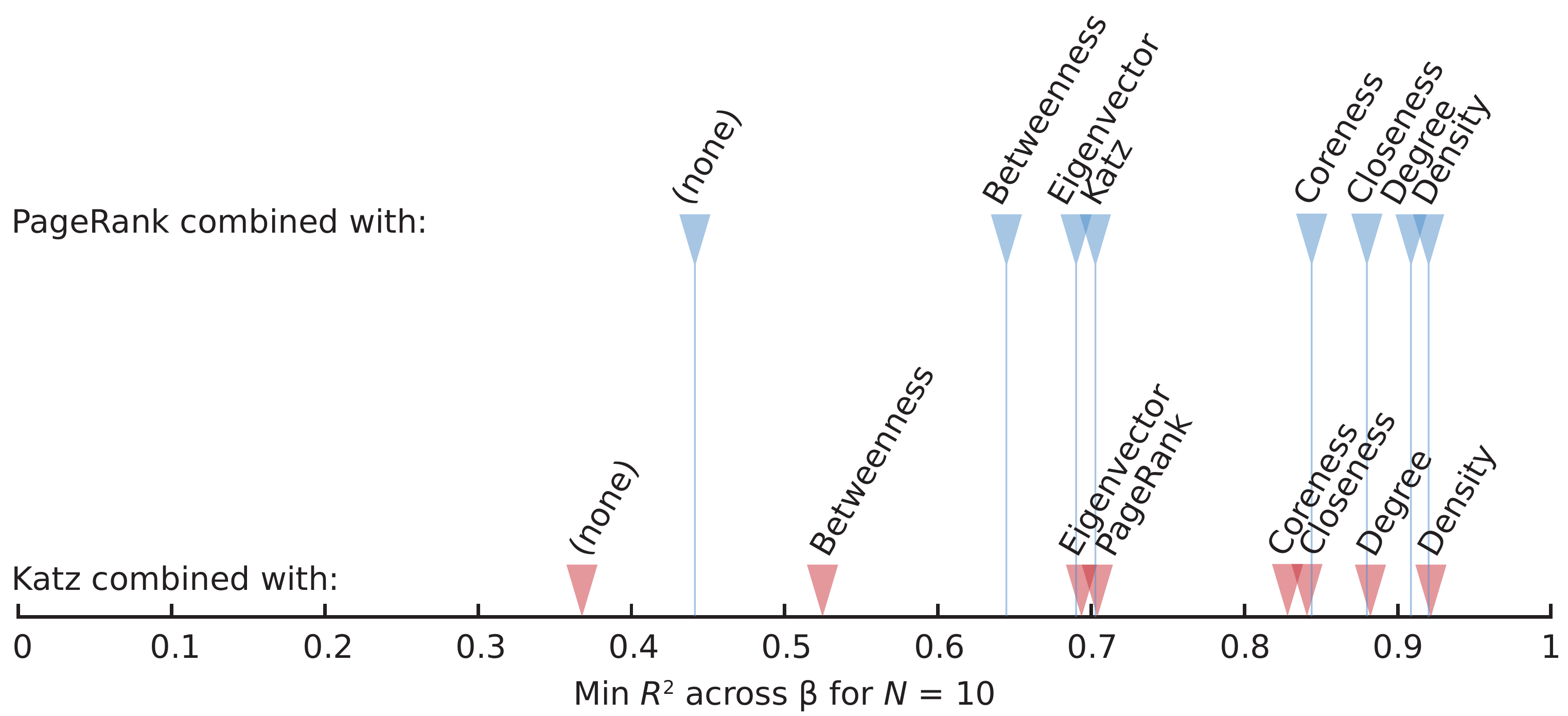}
 \caption{{\bf Combinations between PageRank or Katz centrality and other measures.} The leftmost markers represent single-feature predictions; the rest are combinations with other measures.}
 \label{fig:combinations}
\end{figure}

\subsection*{Predictability with combinations of measures}

We proceed to investigate how adding another feature can increase the predictability of the expected outbreak size. In Fig.~\ref{fig:regression_summary-2centr}, we plot the best performing combinations of two features. A first thing to notice is that, going from one to two features, the $R^2$ values increase considerably. By the intuition given in Figs.~\ref{fig:eigen_vs_omega_scatterplot_degree} and~\ref{fig:eigen_vs_omega_scatterplot_density}, certain structural measures can complement each other to a great extent. Only for very large $\beta$, $R^2$ drops below 0.9. Second, we notice that closeness centrality---the best one-feature predictor---is now overtaken in performance. For predictions with two features, the combination of degree with any spectral centralities performs well. This means that the spectral centralities, although not performing well by themselves, complement degree for larger $\beta$ (while for smaller $\beta$, degree performs well by itself).

The most fundamental information we have not included in the prediction so far is the number of edges in the graph. That is a different type of feature in that it is the same for all nodes in the same graphs, and only a tool to distinguish nodes in graphs of different edge densities. In Fig.~\ref{fig:regression_summary-density-1centr}, we show the coefficient of determination of one of our network centralities in combination with the edge density of the graph the node belongs to. By comparing to Fig.~\ref{fig:regression_summary-2centr}, we can see that the predictive performance is comparable to the case of two  network centralities. The two top-scoring combinations of Fig.~\ref{fig:regression_summary-2centr}---degree and PageRank, and degree and Katz, respectively---are replaced by PageRank and Katz together with the density. Thus, roughly speaking, the graph density adds equally useful information as the degree; and of course, if a small graph has many edges, then many of its nodes have relatively large degrees.

For a further analysis of how different centralities complement one another, in Fig.~\ref{fig:combinations} we display the $R^2$ values of PageRank and Katz in combination with all the others. This shows the observation above more clearly---degree adds similar information as density, and the size-sensitive centralities complement PageRank and Katz better than, e.g., betweenness. The rationales of Katz and PageRank are similar, and so is most of their behavior combined with other measures. Betweenness and degree, however, stand out as improving PageRank much more than Katz. Furthermore, closeness improves PageRank more than degree does, but the difference is small.

\begin{figure}
\begin{adjustwidth}{-2.1in}{0in}
 \includegraphics[width=1.4\textwidth]{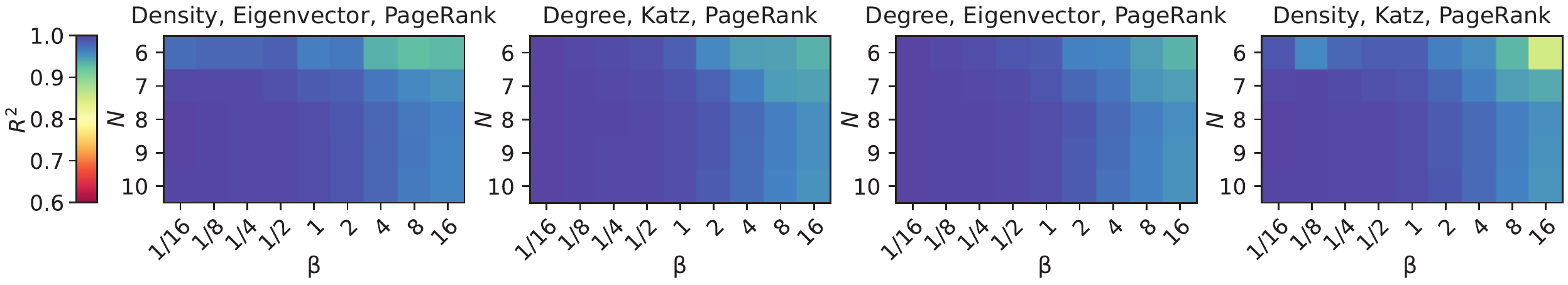}
 
 \caption{{\bf The most predictive triplets of node centralities (including the number of edges).} The coefficient of determination $R^2$ when $\Omega(\beta)$ is estimated over graphs of $N$ nodes. The centrality triplets appear in decreasing order of the largest minimum $R^2$ across $\beta$ values at $N=10$: these four combinations reach $R^2$ values between 0.96 and 0.95 (and many other triplets, not shown here, also score above $R^2=0.90$).}
 \label{fig:regression_summary-3centr}

\end{adjustwidth}
\end{figure}

In Fig.~\ref{fig:regression_summary-3centr}, we extend our investigation to three features. This time we do not separate the edge density from the other features. We show the combinations whose lowest coefficients of determination at $N=10$ are as high as possible. For three features, $R^2$ approaches one---for the combination edge density, eigenvector centrality and PageRank, the least well predicted $\beta,N$-pair has an $R^2$ as large as $0.960$. Including even more features does not give a dramatic improvement in the performance. The situation is similar to the two-feature case in the sense that the spectral centralities are doing better at the expense of, e.g., closeness and coreness.

\begin{figure}
 \includegraphics[width=0.98\textwidth]{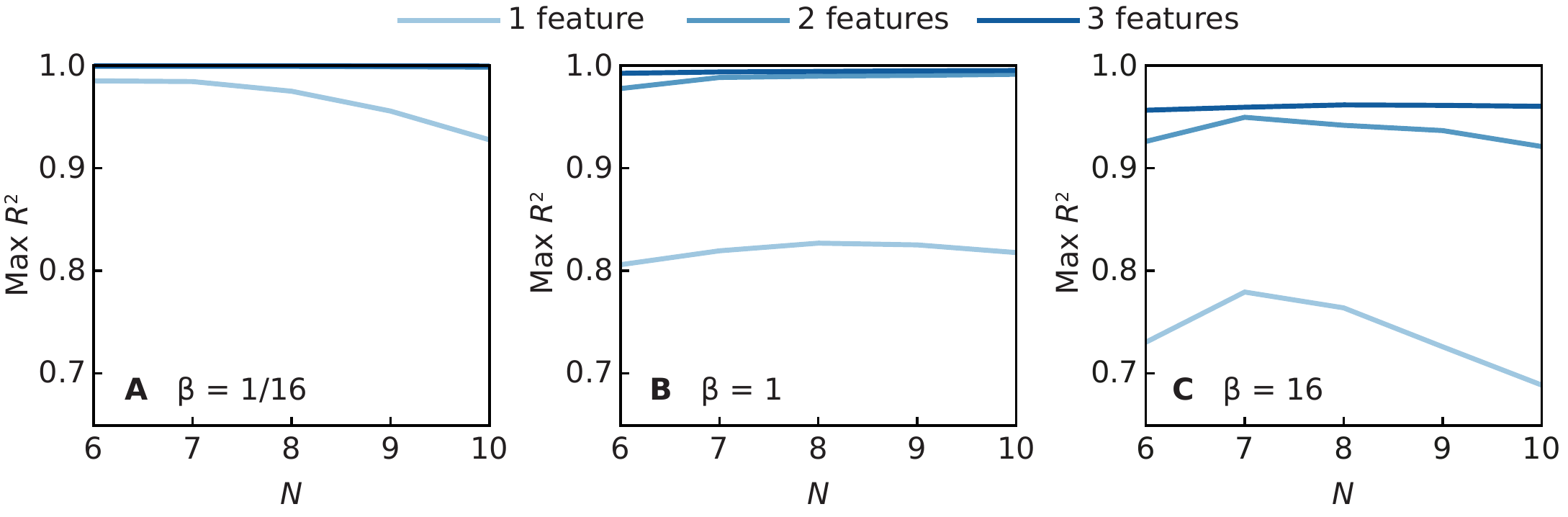}
 \caption{{\bf Size scaling of the best predictability for one, two and three features.} The three panels represent less (A), medium (B) and more (C) contagious diseases}
 \label{fig:R2N_overview}
\end{figure}

Unlike the case of only one feature, when the number of features is two or more, the predictability among the best combinations of predictors is consistently worse for large $\beta$ values. This observation (in agreement with Ref.~\cite{scarpino2019predictability}) means that there are network structures not captured by any of our eight features that affect $\Omega$ in this region; and what that would be, we have to leave as a question for the future. Note that as $\beta$ increases, the range of $\Omega$ decreases, so, in absolute terms, the network structure matters less. If we were relying on stochastic simulations this could potentially be an explanation (fluctuations would affect $R^2$ more), but we do use exact values of $\Omega$.

Besides decreasing with $\beta$, the predictability also decreases somewhat with the size of the network. However, the larger the number of features used as predictors, the more stable the prediction performance is. We highlight this in Fig.~\ref{fig:R2N_overview} where we plot the highest $R^2$ values over all configurations of one, two or three features; with 3 features, the performance score remains stable in this interval of network sizes. As mentioned before, unlike the majority of the literature, we are not primarily interested in the $N\to\infty$ limit. This result is interesting for a basic understanding of the predictability of dynamical systems on networks as one intuitively would think that the larger fluctuations in small networks would make them less predictable. If one considers a specific network model, we believe predictability would increase with system size.

\subsection*{Prediction maps}

\begin{figure}[tb]
 \includegraphics[width=.95\textwidth]{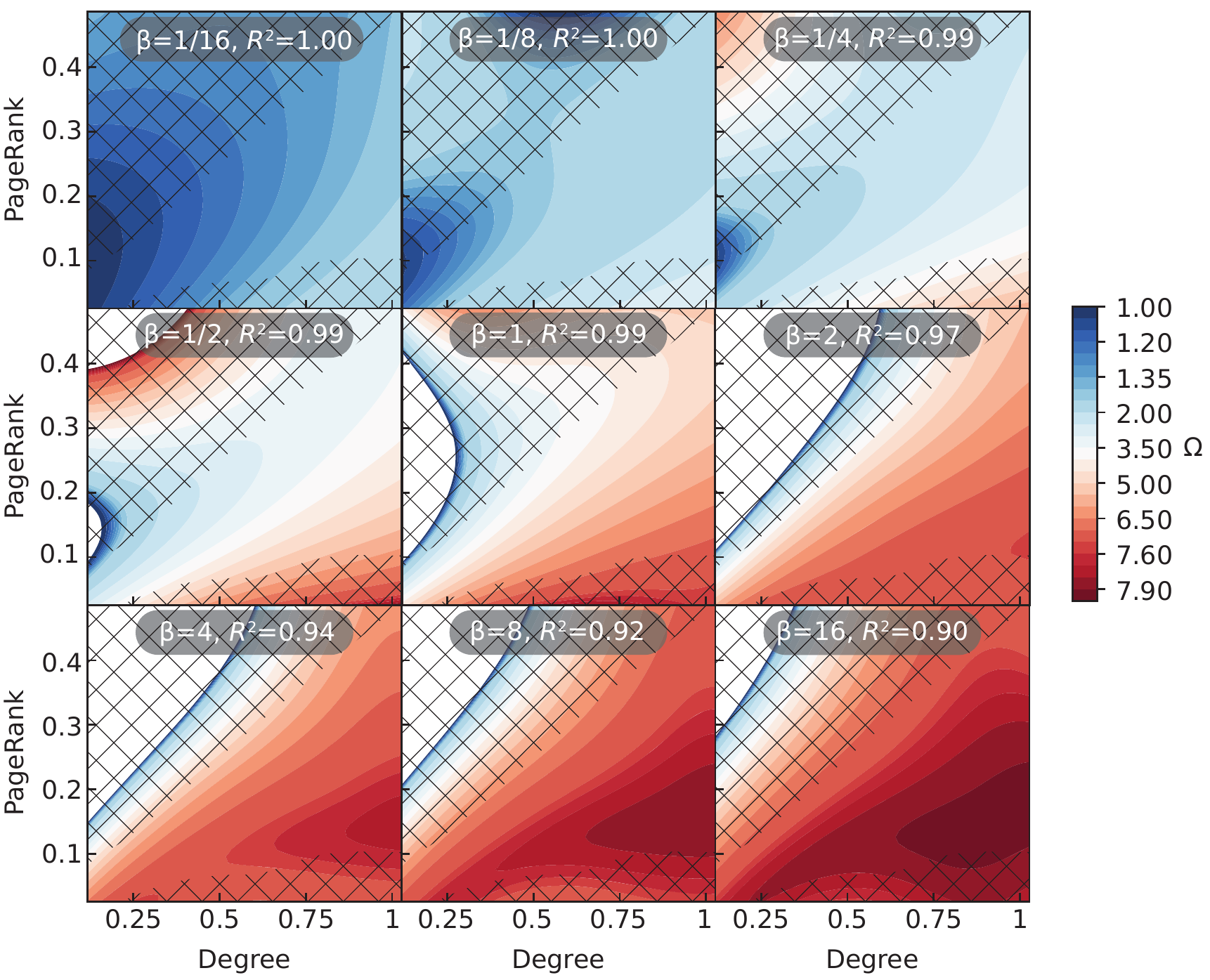}

 \caption{{\bf Prediction maps for the combination of degree and PageRank at different transmission rates.} In each of these subpanels for nine infection rates $\beta$, the degree centrality is given on the x-axis and PageRank on the y-axis. The diagonal grid shows regions where no real graph exists. $N=8$ for all panels.}
 \label{fig:regression-models-Degree-PgRank}
\end{figure}

In our final analysis, we look closer at the statistical models that we learned. The models are visualized in their entirety (see the Methods section). In Fig.~\ref{fig:regression-models-Degree-PgRank}, we show the prediction of outbreak sizes by the best performing combination of two features (the degree and PageRank centralities). Since the regressors see the features as continuous, this type of plot forms a continuous map of $\Omega$ in the parameter space. The real values of all quantities we plot will not fill out the space, but rather form a pattern of points. In the figures, regions of parameter space devoid of data points are marked by a diagonal grid. 

Even if it is meaningless to talk about predictions at coordinates other that graphs can actually attain, these continuous prediction maps visually express the joint contribution of the quantities better than any plot containing only the valid points. Reading the plot by increasing $\beta$ values gives a dynamic sense of the shifting roles of the two features.

In Fig.~\ref{fig:regression-models-Degree-PgRank}, we show the predicted $\Omega$ for all PageRank and degree values. We can see that the nodes with the highest $\Omega$, for all $\beta$-values, tend to have large degree and low PageRank. Intuitively, one would expect nodes with higher PageRank to perform better than those with lower PageRank~\cite{miller:vacc}. The reason for this counter-intuitive result is that PageRank is normalized per graph, and thus less sensitive to the graph size (compared to e.g.\ degree that is bounded by $N$, or closeness that is bounded by the reciprocal diameter and typically going to zero as $1/\log N$ in network models). This means that a node with high degree and low PageRank is typically a node in a very dense graph, where the other nodes will help propagate the outbreak and thus promote a higher $\Omega$. This illustrates how combining two measures can encode different levels of information---about the local position of a node, and about the entire network’s propensity to sustain epidemic outbreaks. This reasoning also applies to Katz centrality and its combinations (even though Katz centrality is normalized in a different way).

\begin{figure}[tb]
 \includegraphics[width=.95\textwidth]{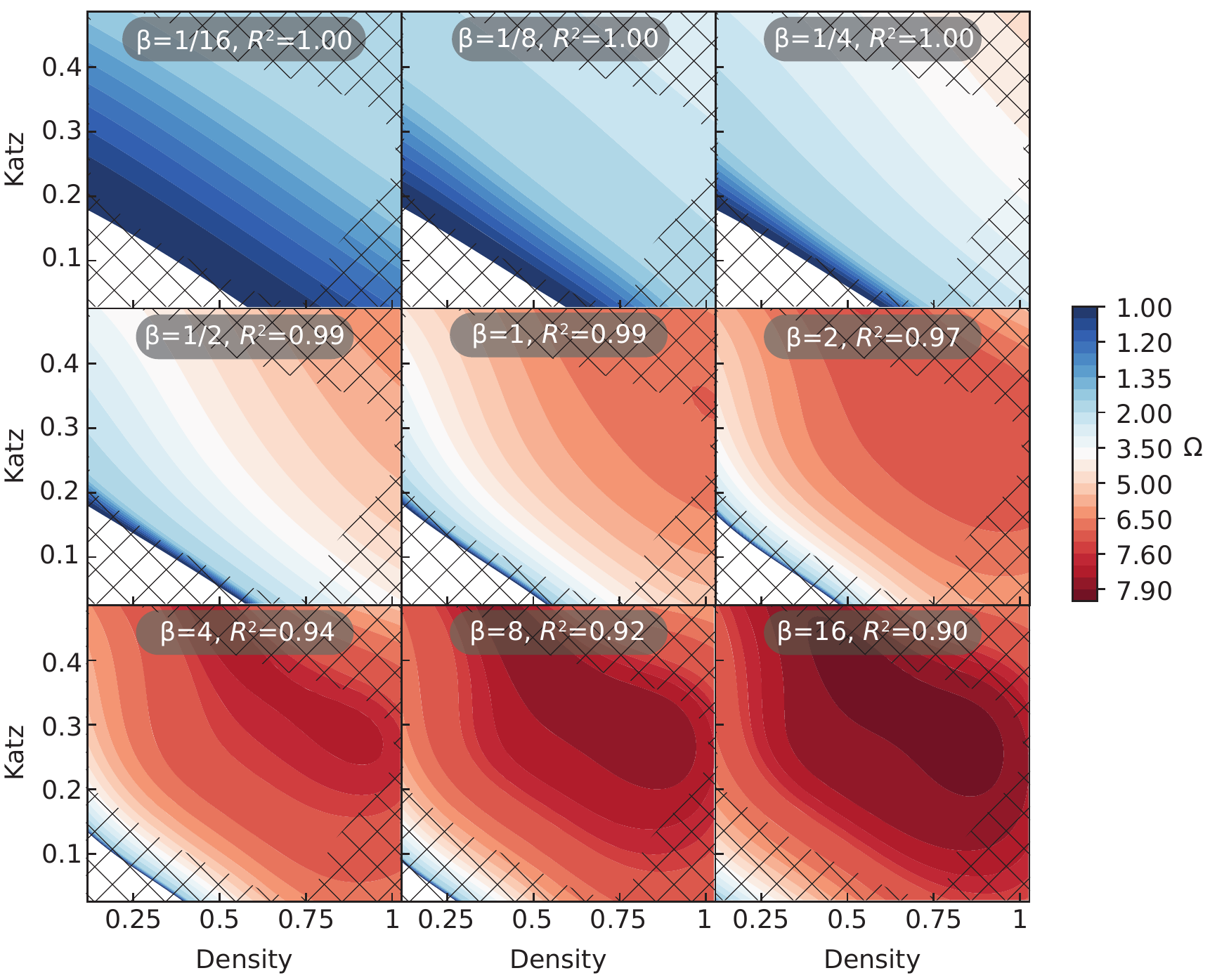}

 \caption{{\bf Prediction maps for the combination of Katz centrality and density at different transmission rates.} This figure corresponds to Fig.~\ref{fig:regression-models-Degree-PgRank} but is for Katz centrality and density instead of degree and PageRank.}
 \label{fig:regression-models-Density-Katz}
\end{figure}

Figure~\ref{fig:regression-models-Density-Katz} shows a plot corresponding to Fig.~\ref{fig:regression-models-Degree-PgRank} but involving the combination of edge density and Katz centrality, which performs equally well as the combination of edge density and PageRank. In this case, $\Omega$ increases with both features and the prediction map changes more smoothly than for the PageRank and degree system. Nodes in networks of medium and high density are more likely to have a larger influence, but the value of the Katz centrality is also discriminative: while all nodes from a very dense network will seed a large outbreak, not any node from an average-density network will also do so, but only those with a maximum value for their Katz centrality. 

\begin{figure}[tb]
 \includegraphics[width=.85\textwidth]{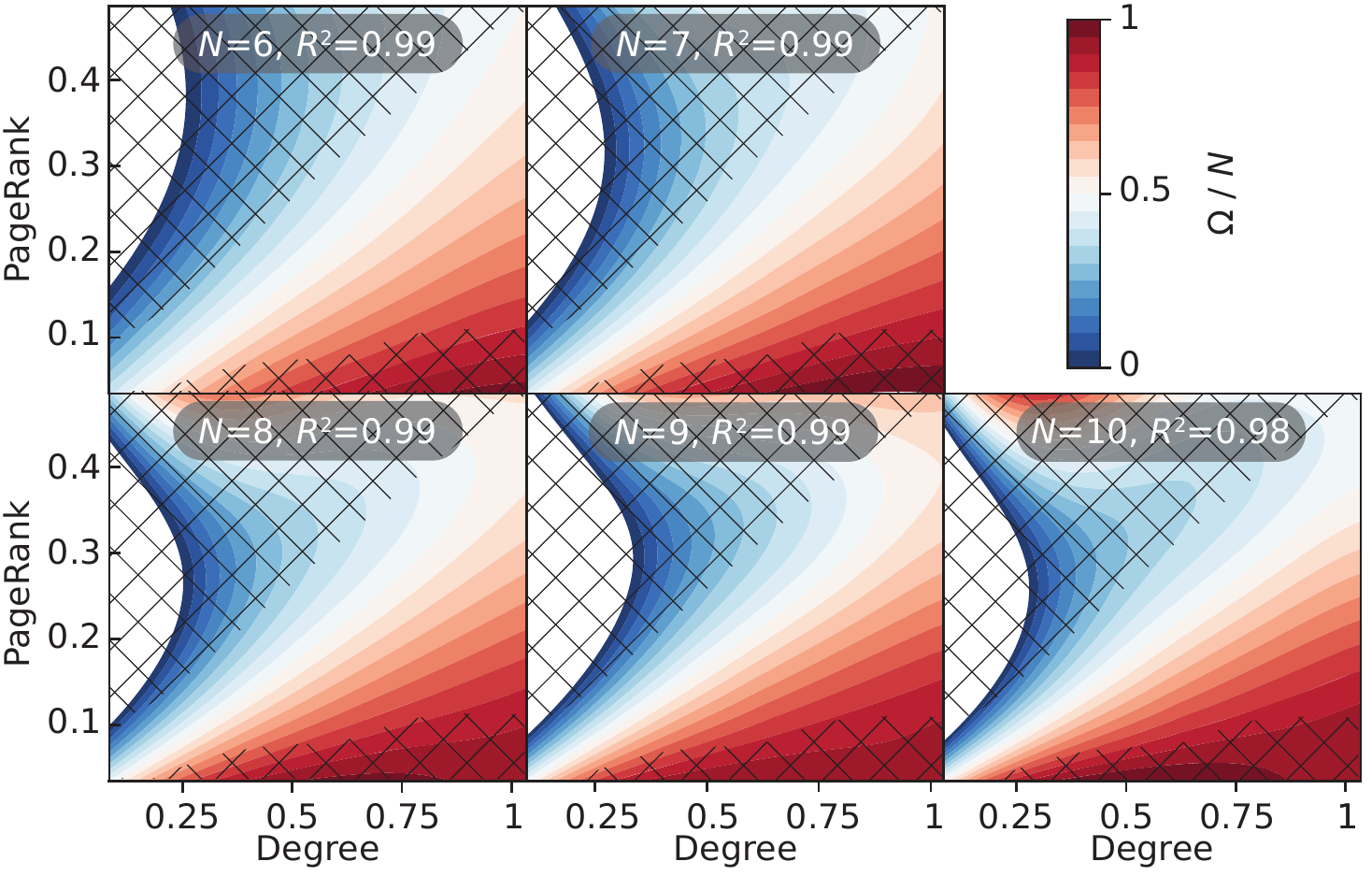}

 \caption{{\bf Prediction maps for the combination of PageRank and degree centrality for different graph sizes.} This figure corresponds to Fig.~\ref{fig:regression-models-Degree-PgRank} but the transmission rate is fixed to $\beta=1$ and we vary the system size. To be able to compare different systems sizes, we plot $\Omega/N$ rather than $\Omega$.}
 \label{fig:regression-models-across-N}

\end{figure}

In our final analysis, in Fig.~\ref{fig:regression-models-across-N} we investigate the dependence on $N$ of the prediction maps of PageRank and degree. In this plot we keep $\beta=1$, so that the panel for $\beta=1$ of Fig.~\ref{fig:regression-models-Degree-PgRank} corresponds to the panel for $N=8$. In general, the size effects are small. The change is smooth, so the general picture would probably extrapolate to much larger $N$. 

\section*{Discussion}

In this work, we have addressed the problem of finding important nodes with respect to disease spreading in networks. All other studies we are aware of phrase this as a problem of ranking nodes in a given network and validating against a ranking obtained based on disease-spreading models. We, on the other hand, try to predict the actual value, not the ranking, from the values of standard network-positional measures. As opposed to other studies, we use combinations of these network measures, and statistical learning. A limitation of statistical learning is that it learns better models for feature values where there are sufficient training data points; areas on the periphery of the prediction maps where few examples exist (e.g., there is only one nonisomorphic graph of maximum density in the data set) will be predicted less accurately.

Our work can be directly applicable to designed interaction networks (such as networks of animal trade~\cite{bajardi,dawson2015epidemic}) and situations where social networks can be mapped out sufficiently and are relevant for outbreaks (such as influenza among college students~\cite{christakis2010social}). At a larger scale, when the social ties are no longer possible to record comprehensively, our work needs to be extended by estimators of centralities that rely on local information~\cite{ruggeri,zhang2015estimating}.

Apart from direct applications, our paper sheds light on the fundamental question of how network structure affects the predictability of outbreaks. Since the importance of nodes depends on the network structures very non-linearly, it is a harder prediction task than ranking nodes. Still, the best statistical models we learned (using the seven standard network measures as predictors) are able to reach a worst-case coefficient of determination as high as $R^2=0.69$ with one predictor, $0.92$ with two, and $0.96$ for three predictors. With a single feature, we find the degree centrality the best for very low $\beta$ and closeness the best otherwise. This confirms the findings from~\cite{3faces}, whereas others find degree to be the best for the entire parameter space~\cite{arruda} or only the largest and smallest $\beta$. The most successful combinations of features typically involve one normalized spectral centrality, such as PageRank or Katz centrality, and one measure sensitive to the edge density in the graphs (such as edge density itself, or degree).

There are many directions worth exploring at the interface between machine learning and theoretical epidemiology, more or less similar to the current work~\cite{hay2013big}. Straightforward continuations would be to investigate: larger, model networks by stochastic simulations; newer, more specialized measures for predicting epidemic importance~\cite{vsikic2013epidemic}; or other scenarios to optimize (such as targeted vaccination~\cite{miller:vacc,rushmore} or sentinel surveillance~\cite{3faces,christakis2010social}). One question remaining is why the prediction using multiple features is worse at high $\beta$. This means that there are network structures not captured by any of our eight measures that affect the importance---is there some simple, undiscovered network measure capturing these?

\section*{Acknowledgments}

We thank the organizers of the YEP 2019 workshop on Information Diffusion on Random Networks at TU Eindhoven, where we initiated this work.

\nolinenumbers

\end{document}